\documentclass[aps,prb,twocolumn,groupedaddress]{revtex4-1}
\usepackage{graphicx}
\usepackage{dcolumn}
\usepackage{bm}
\usepackage{amsmath}
\usepackage{amssymb}
\usepackage{color}

\begin{document}

\title{Blinking membrane patterns induced by protein binding/unbinding}

\author{Hiroshi Noguchi}
\email[]{noguchi@issp.u-tokyo.ac.jp}
\affiliation{Institute for Solid State Physics, University of Tokyo, Kashiwa, Chiba 277-8581, Japan}


\begin{abstract}
Nonequilibrium membrane pattern formation is studied using meshless membrane simulation.
Bound proteins are considered to have two states that generate different membrane spontaneous curvatures.
Protein binding and unbinding occur cyclically owing to chemical potential differences,
as  an off-lattice active Potts model.
It is found that this cyclic binding/unbinding can induce blinking domains, with oscillating size: 
convex domains of the proteins with a higher spontaneous curvature grow,
and subsequently, the proteins change to the other state with a lower spontaneous curvature,
resulting in domain shrinkage. These processes repeat.
In thermal equilibrium, hexagonal convex domains are formed by the competition between bending and surface tension energies, so that they are stably formed only under positive surface tension.
However, blinking domains can form even in tensionless membranes.
\end{abstract}
\maketitle

\section{Introduction}

Various types of spatiotemporal patterns, including traveling and standing waves, have been observed in living cells and tissues.\cite{meri21,wu21,beta17,nogu24c,verg24}
Traveling waves can induce membrane shape oscillations through the periodic binding of curvature-inducing proteins.\cite{lits18,wu18}
Membrane shape transformation is a crucial process in various biological functions, such as cell transport, motility, and division.\cite{mcma05,zimm06,baum11,suet14,kaks18,beth18,svit18,lutk12,sack21,li25} 
During vesicle transport, the assembly of curvature-inducing proteins, such as clathrin and coat protein complexes,
generates spherical buds.\cite{kaks18,beth18,joha15,bran13,hurl10,mcma11}
During the locomotion of amoeboid cells, excitable waves on membranes play a key role in changing cell morphology.\cite{igle26,arai10,tani13,huan13}
The position of cell division during mitosis is determined by the pole-to-pole standing wave of Min proteins on plasma membranes.\cite{lits18,lutk12,taka22,ren25} 
Further, the cell polarity of eukaryotic embryos is determined by a PAR protein pattern.\cite{meri21,lutk12,hoeg13}

The binding of curvature-inducing proteins to membranes at thermal equilibrium has been intensively studied through experimental, theoretical, and computational approaches.\cite{baum11,has21,tsai21,nogu25c,sade21,vida26}
These proteins can sense and generate membrane curvature.
The curvature dependences of these proteins have been reproduced using mean-field theories.\cite{nogu25c,nogu21a,nogu22a,nogu23b,nogu24} 
The assembly of these proteins produces spherical buds and cylindrical membrane tubes\cite{baum11,mim12a,fros08,adam15,rama18,nogu16,nogu22b} as well as periodic patterns such as hexagonal\cite{gout21} and checkerboard\cite{nogu23} arrays of curved domains and beaded-necklace-like membrane tubes.\cite{nogu21b}

In contrast to equilibrium binding, nonequilibrium binding remains much less explored.
Coupling of membrane deformation and reaction-diffusion dynamics has been simulated using a dynamically triangulated membrane model.\cite{tame20,tame21,tame22,nogu23a}
Traveling waves can be enhanced or suppressed by membrane deformation.\cite{tame21,tame22,nogu23a}
Reflection-symmetric shape deformation can stabilize Turing patterns.\cite{tame20}
However, by changing a flat surface to a reflection-asymmetric surface,
static Turing patterns can change to propagating or chaotic patterns  even without dynamic surface deformation.\cite{nish22,nish25}
However, the effects of thermal fluctuations on the spatiotemporal patterns of deformable membranes at the molecular scale have not been investigated so far.
Previously, we simulated the membrane spatiotemporal patterns resulting from
the binding and unbinding of laterally isotropic molecules at both membrane surfaces,
in which the binding of each molecule is treated as a discrete process.\cite{nogu25a}
Biphasic domains move ballistically or diffusively depending on the conditions,
and aligned domains exhibit time-irreversible fluctuations.

\begin{figure}[t]
\includegraphics{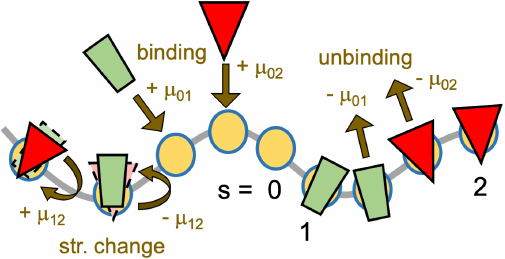}
\caption{
Schematic of the binding and unbinding of curvature-inducing proteins with two states.
The gray curve and yellow circles represent the membrane and its binding sites, respectively.
The proteins have two shapes (green trapezoids and red triangles)
and bind to the membrane with the  binding chemical potentials $\mu_{01}$ and $\mu_{02}$, respectively.
The unbound state is $s=0$, and the bound states are $s=1$ and $2$ 
with spontaneous curvatures $C_0=0.05$ and $0.1$, respectively.
The chemical potential from  $s=1$ to $s=2$ is $\mu_{12}$.
}
\label{fig:cart}
\end{figure}

In this study, we investigated the spatiotemporal domain patterns caused by cyclic protein binding to one side of the membranes.
The proteins have two states with different isotropic spontaneous curvatures of the same sign (see Fig.~\ref{fig:cart}).
We used a meshless membrane model with three states (one unbound and two bound) as an off-lattice version of an active Potts model.\cite{nogu25a,nogu24a,nogu24b,nogu25b,nogu26a,nogu26c} 
This membrane model can be used to simulate large membrane deformation, including topological changes, and the membrane properties (bending rigidity, spontaneous curvature, and edge line tension) can vary widely in a fluid phase.
We observed blinking domain (BD) patterns, wherein the domain size cyclically changes without ballistic motion.
Such standing-wave-like patterns have not been obtained in either lattice\cite{nogu24a,nogu24b,nogu25b,nogu26a,nogu26c} or off-lattice\cite{nogu25a}
active Potts models.
The membrane bending keeps the domain location unchanged.

The simulation model and method are described in Sec.~\ref{sec:method}.
The results are presented and discussed in Sec.~\ref{sec:results}.
Section~\ref{sec:eq} briefly describes the phase behaviors in thermal equilibrium as reference states.
Sections.~\ref{sec:t1} and \ref{sec:t0} describe nonequilibrium pattern formations at high and low surface tensions, respectively.
Finally, a summary is presented in Sec.~\ref{sec:sum}.

\section{Simulation Model and Method}\label{sec:method}

In a meshless membrane model,
membrane particles self-assemble into a single-layer sheet in the fluid phase.\cite{nogu09,nogu06,shib11,nogu19}
Proteins or other molecules bind to and unbind from the membrane.
Here, each membrane particle is a binding site with three states: one unbound state (the state $s=0$) and two bound states ($s=1$ and $2$), as shown in Fig.~\ref{fig:cart}.
These two bound states are laterally isotropic and have different spontaneous curvatures with the same sign.
The chemical potentials binding to the two bound states are  $\mu_{01}$ and  $\mu_{02}$,
and the chemical potential from the bound state $s=1$  to $s=2$ is  $\mu_{12}$.
Therefore, in thermal equilibrium,  $\mu_{02} = \mu_{01} + \mu_{12}$.
However, out of equilibrium,
owing to energy inputs such as ATP hydrolysis and  concentration deviation from the equilibrium relation, $\mu_{02} \ne \mu_{01} + \mu_{12}$.

The position and orientation of the $i$-th particle are ${\bm{r}}_{i}$ and ${\bm{u}}_i$, respectively.
The membrane particles interact with each other through a potential $U=U_{\mathrm {rep}}+U_{\mathrm {att}}+U_{\mathrm {bend}}+U_{\mathrm {tilt}}+U_{\mathrm {pp}}$.
The potential $U_{\mathrm {rep}}$ is an excluded-volume interaction with diameter $\sigma$ for all particle pairs.
The solvent is implicitly accounted for by an effective attractive potential  as follows:
\begin{equation} \label{eq:U_att}
\frac{U_{\mathrm {att}}}{k_{\mathrm{B}}T} =  \frac{\varepsilon_{\mathrm{att}}}{4}\sum_{i} \ln[1+\exp\{-4(\rho_i-\rho^*)\}],
\end{equation}
where  $\rho_i= \sum_{j \ne i} f_{\mathrm {cut}}(r_{i,j})$,
 $\rho^*$ is the characteristic density with $\rho^*=7$, $\varepsilon_{\mathrm{att}}=8$,
and $k_{\mathrm{B}}T$ is the thermal energy, as described in our previous studies.\cite{gout21,nogu23,nogu25a}
$f_{\mathrm {cut}}(r)$ is a $C^{\infty}$ cutoff function\cite{nogu06}
 and $r_{i,j}=|{\bf r}_{i,j}|$, with ${\bf r}_{i,j}={\bf r}_{i}-{\bf r}_j$:
\begin{equation} \label{eq:cutoff}
f_{\mathrm {cut}}(r)=
\exp\Big\{b\Big(1+\frac{1}{(r/r_{\mathrm {cut}})^n -1}\Big)\Big\}\Theta(r_{\mathrm {cut}}-r),
\end{equation}
where $\Theta(x)$ is the unit step function, 
 $n=6$, $b=\ln(2) \{(r_{\mathrm {cut}}/r_{\mathrm {att}})^n-1\}$,
$r_{\mathrm {att}}= 1.9\sigma$, and $r_{\mathrm {cut}}=2.4\sigma$.
The bending and tilt potentials
are represented as
\begin{eqnarray} \label{eq:ubend}
\frac{U_{\mathrm {bend}}}{k_{\mathrm{B}}T} &=& \frac{k_{\mathrm {bend}}}{2} \sum_{i<j} ({\bm{u}}_{i} - {\bm{u}}_{j} - C_{\mathrm {bd}} \hat{\bm{r}}_{i,j} )^2 w_{\mathrm {cv}}(r_{i,j}), \\
\frac{U_{\mathrm {tilt}}}{k_{\mathrm{B}}T} &=& \frac{k_{\mathrm{tilt}}}{2} \sum_{i<j} [ ( {\bm{u}}_{i}\cdot \hat{\bm{r}}_{i,j})^2
 + ({\bm{u}}_{j}\cdot \hat{\bm{r}}_{i,j})^2  ] w_{\mathrm {cv}}(r_{i,j}),\  
\end{eqnarray}
where 
 $\hat{\bm{r}}_{i,j}={\bm{r}}_{i,j}/r_{i,j}$ and $w_{\mathrm {cv}}(r_{i,j})$ is a weight function. 
The spontaneous curvature is given by $C_0 = C_{\mathrm {bd}}/2\sigma$.\cite{shib11} 

\begin{figure}[t]
\includegraphics[]{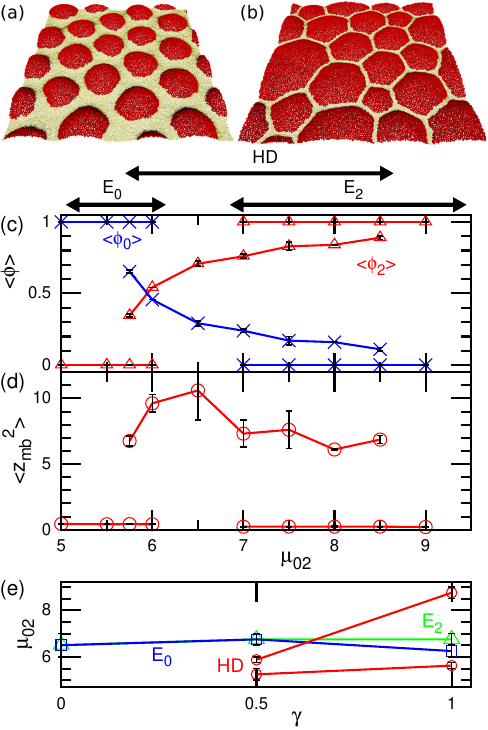}
\caption{
Binding of proteins of the $s=2$ state under thermal equilibrium.
(a)--(b) Snapshots of the HD phase for (a) $\mu_{02}=6$ and (b) $\mu_{02}=8$ at the surface tension $\gamma = 1$.
Red (dark gray) and yellow (light gray) spheres represent the bound and unbound membrane particles ($s=2$ and $s=0$), respectively.
The $s=2$ convex domains are hexagonally aligned. 
(c)--(d) Dependence on $\mu_{02}$ at $\gamma = 1$.
(c) Mean density $\langle\phi\rangle$ of the $s=0$ and $s=2$ states.
(d) Mean membrane vertical span $\langle{z_{\mathrm {mb}}}^2\rangle$.
The bidirectional arrows at the top of (c) represent the ranges of the membrane phases.
(e) Phase diagram for $\gamma$ vs. $\mu_{02}$.
The blue squares and green triangles represent
the upper and lower limits of the $s=0$ and $s=2$ dominant phases (E$_0$ and E$_2$), respectively.
The HD phase lies between the two red lines marked with circles.
}
\label{fig:eq}
\end{figure}

The repulsive interactions of different states are given by 
\begin{equation} \label{eq:upp}
\frac{U_{\mathrm {pp}}}{k_{\mathrm{B}}T} =  \sum_{s_i\ne s_j} \varepsilon_{\mathrm{pp}} \exp\big( \frac{1}{r_{ij}/r_{\mathrm {cut}} -1} + b_0 \big) ,
\end{equation}
where $b_0= r_{\mathrm {cut}}/(r_{\mathrm {cut}}-\sigma)$.\cite{nogu12a}
At $r_{ij}=\sigma$, the potential height is $\varepsilon_{\mathrm {pp}}k_{\mathrm{B}}T$.
In this study, $\varepsilon_{\mathrm {pp}}=2$ is used to generate phase separation.

The position ${\bf r}_{i}$ and 
 orientation ${\bf u}_{i}$ of the membrane particles are updated through underdamped Langevin equations,
integrated using the leapfrog algorithm\cite{alle87,nogu11}
with $\Delta t=0.002\tau$. The time unit is $\tau= \sigma^2/D_0$,
where $D_0$ is the diffusion coefficient of free membrane particles.
A membrane particle is randomly selected, and
the membrane state is stochastically switched from $s=k$ to $[k-1]$ or $[k+1]$ with $1/2$ probability  using a Metropolis Monte Carlo (MC) procedure with the acceptance rate $p_{\mathrm {acpt}}$, where $[k]$ indicates $k \bmod q$:
\begin{equation}\label{eq:Metro}
p_{\mathrm {acpt}} = \mathrm{min} \bigg[1, \exp\Big(- \frac{\Delta H}{k_{\mathrm{B}}T}\Big) \bigg].
\end{equation}
Here, $\Delta H= \Delta U - \mu_{\alpha\beta}$, where $\Delta U = U_{\mathrm{new}} - U_{\mathrm{old}}$ is the energy difference between the new and old states,
and $\mu_{\alpha\beta}$ is the chemical potential between $s=\alpha$ and $\beta$, and  $\mu_{\beta\alpha}=-\mu_{\alpha\beta}$.
The state flips are performed by one MC trial per particle at every MC step of $\tau_{\mathrm {MC}}=0.01\tau$.
Hereafter, we use $k_{\mathrm{B}}T=1$ as the energy unit, particle diameter $\sigma=1$ as the length unit,
and time unit $\tau=1$ for simplicity.
The unit length is considered to be the size of a binding site, which varies from $\sigma=5$ to $100$\,nm
depending on the type of binding molecules.

A membrane comprising $25,600$ membrane particles is simulated under periodic boundary conditions.
The $N\gamma T$ ensemble is used, where $N$ is the total number of particles and $\gamma$ is the mechanical surface tension.\cite{fell95,nogu12}
We primarily use a tension of $\gamma=1$, which is sufficiently large to prevent the budding of a bound domain.\cite{gout21,nogu23}
This tension corresponds to an average tension $\approx  0.04$\,mN/m, which is well below the usual lysis tension 
($1$--$25$\,mN/m)\cite{evan00,evan03,ly04} for $\sigma\approx 10$\,nm.

For unbound particles,
$C_0=0$ and bending rigidity $\kappa_{\mathrm {u}}/k_{\mathrm{B}}T=16.1 \pm 1$ ($k_{\mathrm {bend}}=k_{\mathrm{tilt}}=10$) are used.
These are typical values for lipid membranes.\cite{kara23,dimo14,hu12}
The two bound states ($s=1$ and $2$) have spontaneous curvatures $C_0= 0.05$ and $0.1$, respectively (Fig.~\ref{fig:cart}),
with a high bending rigidity $\kappa_{\mathrm {b}}/k_{\mathrm{B}}T=144 \pm 7$ ($k_{\mathrm {bend}}=k_{\mathrm{tilt}}=80$).
In this meshless membrane model, the saddle-splay modulus $\bar{\kappa}$ is proportional to $\kappa$, as $\bar{\kappa}/\kappa=-0.9\pm 0.1$.\cite{nogu19} Hence, $\bar{\kappa}$ also changes with protein binding.
Note that protein binding depends on the local Gaussian curvature when $\bar{\kappa}$ is varied by the binding.\cite{nogu25c,nogu22a,nogu21a}
This parameter set is identical to that used in our previous study,\cite{nogu25a} except for the spontaneous curvatures 
(positive and negative spontaneous curvatures are used for the bound states\cite{nogu25a}).

The mean cluster sizes are calculated to characterize various phases. 
Two membrane particles of each state are considered to belong to the same cluster
when the inter-particle distance
is less than $r_{\mathrm {att}}$. 
The mean size of the clusters is 
$N_{s,\mathrm {cl}}= (\sum_{i_{s,{\mathrm {cl}}}=1}^{N_s} i_{s,\mathrm{cl}}^2 n^{s,{\mathrm {cl}}}_i)/N_s$,
where $n^{s,{\mathrm {cl}}}_i$  is the number of clusters with size $i_{s,\mathrm{cl}}$ 
and $N_s$ is the total number of each state.
The mean cluster size of  each state
is normalized by the mean total number as $\chi_{s}=\langle N_{s,\mathrm{cl}}\rangle/\langle N_{s}\rangle$. 
A large percolated cluster results in $\chi \simeq 1$.
The vertical span of the membrane is calculated from 
the membrane height variance as 
${z_{\mathrm {mb}}}^2=\sum_{i}^{N} (z_i-z_{\mathrm G})^2/N$,
where $z_{\mathrm G}=\sum_{i}^{N} z_i/N$. 
The statistical errors are calculated from three or more independent runs
for $10,000\tau$ or longer periods.

\begin{figure}[t]
\includegraphics[]{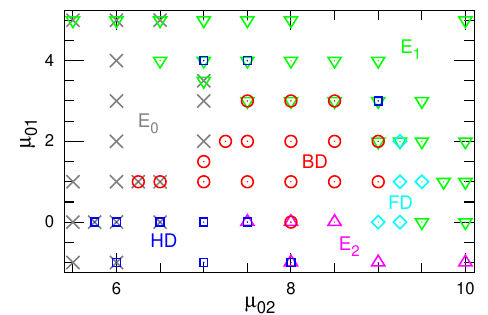}
\caption{
Dynamic phase diagram at $\gamma = 1$.
The gray crosses and green down-pointing and magenta up-pointing triangles represent the homogeneous phases of the states
$s=0$, $1$, and $2$ (E$_0$, E$_1$, and E$_2$), respectively.
The red circles, blue squares, and cyan diamonds represent the BD, HD, and FD patterns, respectively.
The symbols overlap when two modes coexist through hysteresis. 
}
\label{fig:pdt1}
\end{figure}

\section{Simulation Results}\label{sec:results}

\subsection{Equilibrium Phases}\label{sec:eq}

Before simulating the nonequilibrium dynamics,
we investigate the equilibrium phases as reference states.
We simulate the binding and unbinding of the proteins in the $s=2$ state in the absence of the $s=1$ state, using $\mu_{12}=\mu_{02}$ and $\mu_{01}=0$.
At low and high $\mu_{02}$, the unbound and bound states are dominant, and these phases are referred to as E$_0$ and E$_2$, respectively.
Under a sufficiently high surface tension ($\gamma=1$), the bound state forms convex domains with a hexagonal alignment
at intermediate $\mu_{02}$ (Fig.~\ref{fig:eq}), as reported in our previous study\cite{gout21}.
We refer to this hexagonal domain phase as HD. 
As $\mu_{02}$ increases,
the convex domains become larger and change from circular to hexagonal (see Fig.~\ref{fig:eq}(a)--(c)).
 HD displays a large vertical span ${z_{\mathrm {mb}}}^2$, whereas E$_0$ and E$_2$ exhibit small spans (see Fig.~\ref{fig:eq}(d)).
Both the E$_0$--HD and HD--E$_2$ transitions are of the first order, and two phases coexist around the transition points through hysteresis (see Fig.~\ref{fig:eq}(c)).
Although the E$_0$--HD transition is of the second order in the previous study,\cite{gout21}
 adding the repulsive potential $U_{\mathrm {pp}}$ makes it discontinuous.
The repulsion between the unbound ($s=0$) state and bound states ($s=1$ and $2$) are caused by the membrane-mediated Casimir-like interaction\cite{gout21} and this direct repulsion by $U_{\mathrm {pp}}$, whereas
the repulsion between two bound states, $s=1$ and $2$, is caused by the direct repulsion.

In tensionless membranes ($\gamma=0$), the HD phase is unstable. 
The convex domains grow into spherical buds, and the detached buds form vesicles.
The balance of surface tension and bending energy maintains the domains with a spherical-cap shape at $\gamma>0$.\cite{gout21}
Thus, the $\mu_{02}$ range of the HD phase is narrower at $\gamma=0.5$ than that at $\gamma=1$ (see Fig.~\ref{fig:eq}(e)).
Conversely, the boundaries of the E$_0$ and E$_2$ phases are only a little dependent on $\gamma$, since their membranes are flat.

\begin{figure}[t]
\includegraphics[]{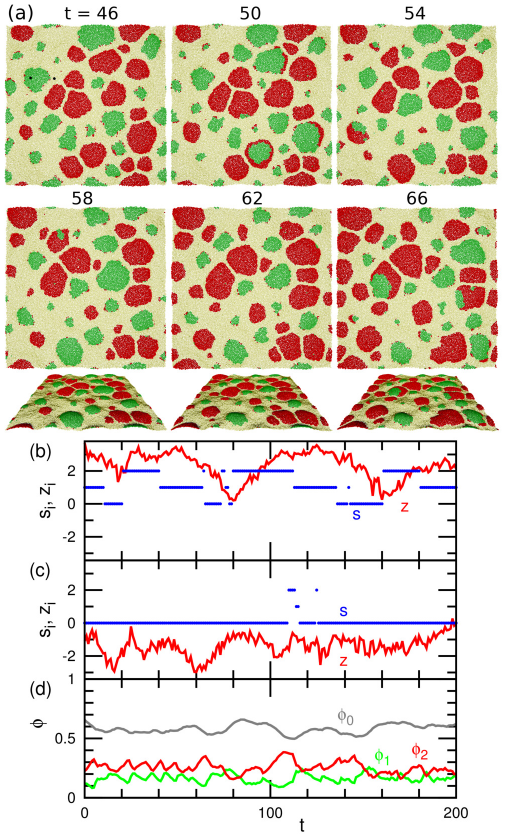}
\caption{
BD pattern at  $\mu_{01} = 2$, $\mu_{02} = 8$, and $\gamma = 1$.
(a) Sequential snapshots. 
Green (medium gray) and red (dark gray) spheres represent the bound membrane particles ($s=1$ and $s=2$), respectively.
Yellow (light gray) spheres represent the unbound particles $s=0$.
Side views are also displayed for the last three snapshots.
Convex $s=2$ (red) domains grow and subsequently change their state to $s=1$ (green), resulting in domain shrinkage.
(b)--(d) Time evolution, corresponding to the snapshots in (a).
(b)--(c)  $s_i$ and $z_i$ of the membrane particle closest to the (b) left and (c) right black filled circles in the snapshot at $t=46$. The origin of the $z$ axis is set at the center of mass of the membrane.
(d) Densities $\phi$ of the three states.
}
\label{fig:BD}
\end{figure}

\begin{figure}[t]
\includegraphics[]{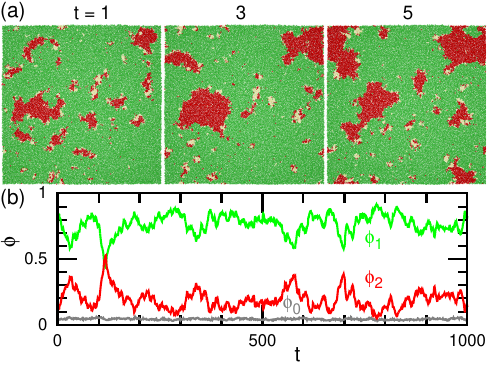}
\caption{
FD pattern at  $\mu_{01} = 0$, $\mu_{02} = 9.25$, and $\gamma = 1$.
(a) Sequential snapshots.
Flat  $s=2$ (red) domains move like amoebas.
(b) Time evolution of the densities $\phi$ of the three states, corresponding to the snapshots in (a). 
}
\label{fig:FD}
\end{figure}

\begin{figure*}[]
\includegraphics[]{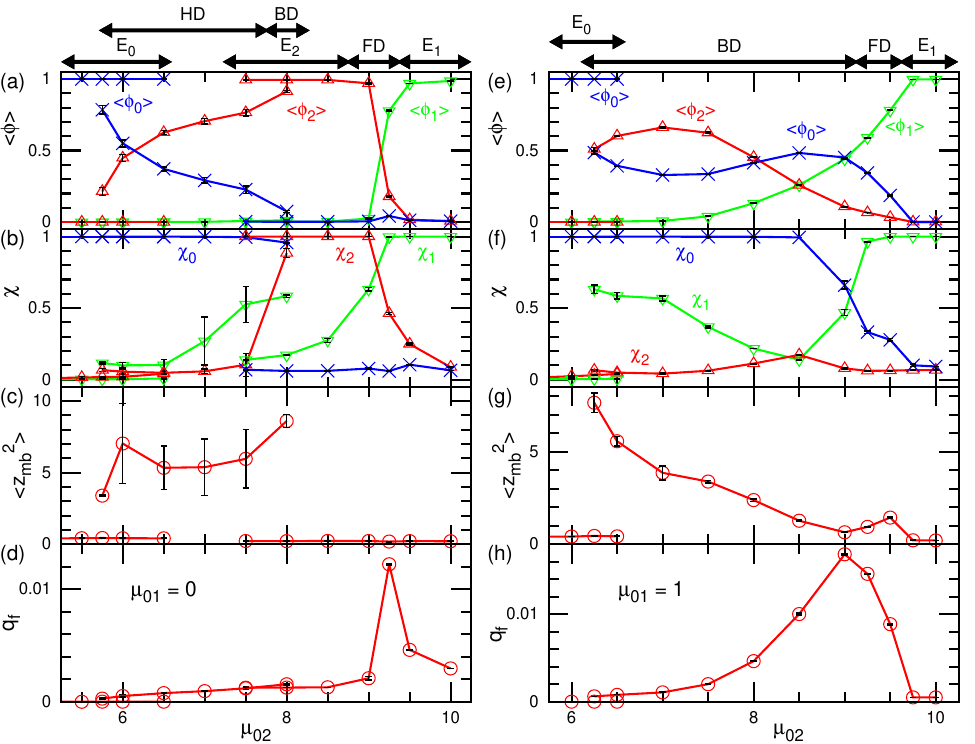}
\caption{
Dependence on $\mu_{02}$ for (a)--(d)  $\mu_{01} = 0$ and (e)--(h) $\mu_{01} = 1$ at $\gamma = 1$.
(a), (e) Mean density $\langle \phi \rangle$ of the three states.
(b), (f) Number ratio $\chi$ belonging to the largest cluster in each state.
(c), (g) Mean membrane vertical span $\langle{z_{\mathrm {mb}}}^2\rangle$.
(d), (h) Mean flipping rate $q_{\mathrm{f}}$ between neighboring states.
The bidirectional arrows at the top of (a) and (e) represent the ranges of the dynamic modes.
}
\label{fig:u0u1}
\end{figure*}

\begin{figure}[t]
\includegraphics[]{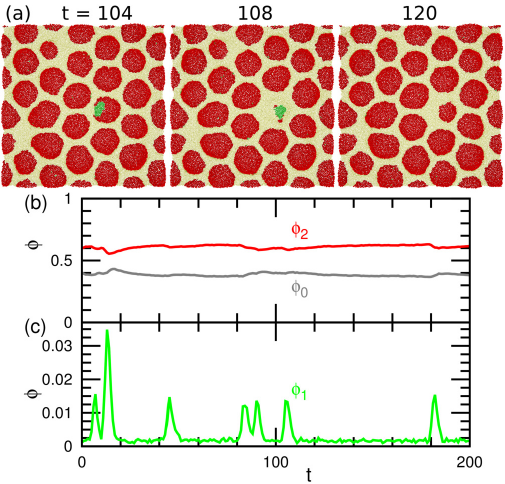}
\caption{
Partially blinking pattern at  $\mu_{01} = 1$, $\mu_{02} = 6.5$, and $\gamma = 1$. 
(a) Sequential snapshots.
Convex $s=2$ (red) domains form a hexagonal alignment,
whereas a few domains exhibit blinking. 
(b)--(c) Time evolution of the densities $\phi$ of the three states, corresponding to the snapshots in (a). 
}
\label{fig:BD2}
\end{figure}

\begin{figure}[t]
\includegraphics[]{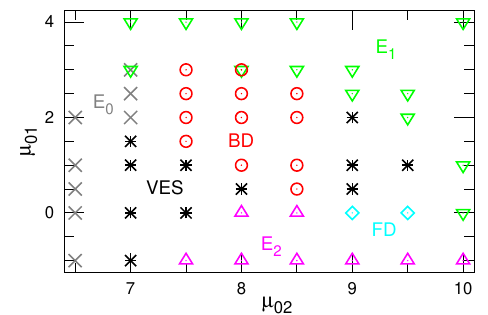}
\caption{
Dynamic phase diagram at $\gamma = 0$.
The gray crosses and green down-pointing and magenta up-pointing triangles represent the homogeneous phases of the states
$s=0$, $1$, and $2$ (E$_0$, E$_1$, and E$_2$), respectively.
The red circles and cyan diamonds represent the BD and FD patterns, respectively.
Black asterisks represent vesicle (VES) formation.
The symbols overlap when two modes coexist through hysteresis. 
}
\label{fig:pdt0}
\end{figure}

\begin{figure}[t]
\includegraphics[]{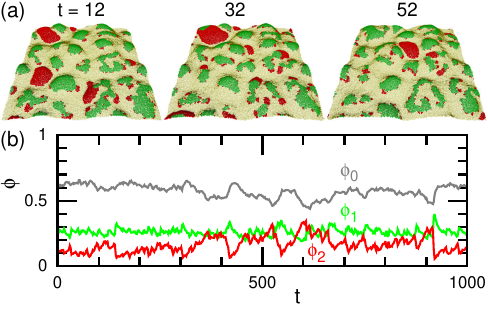}
\caption{
BD pattern at  $\mu_{01} = 1$, $\mu_{02} = 8.5$, and $\gamma = 0$.
(a) Sequential snapshots.
(b) Time evolution of the densities $\phi$ of the three states, corresponding to the snapshots in (a). 
}
\label{fig:BDt0}
\end{figure}

\begin{figure}[t]
\includegraphics[]{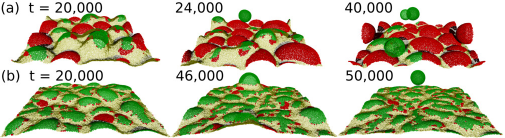}
\caption{
Sequential snapshots of vesicle formation at $\gamma = 0$.
(a) $\mu_{01} = 1$ and $\mu_{02} = 7.5$.
(b) $\mu_{01} = 1$ and $\mu_{02} = 9$.
}
\label{fig:VESt0}
\end{figure}

\subsection{Nonequilibrium Patterns at $\gamma=1$}\label{sec:t1}

In this subsection, we consider a membrane in nonequilibrium ($\mu_{02}-\mu_{01}-\mu_{12}>0$) at $\gamma =1$.
Figure \ref{fig:pdt1} shows a dynamic phase diagram at $\mu_{12}=0$.
Hereafter, we use $\mu_{12}=0$, unless specified otherwise.
The chemical potential imbalance ($\mu_{02}>\mu_{01}$) causes cyclic state flips as $s=0\to 2\to 1\to 0$.
At $\mu_{01}\simeq 2$ and $\mu_{02}\simeq 8$, a novel dynamic mode BD emerges, wherein the sizes of convex domains change cyclically (see Figs.~\ref{fig:BD} and S1 and Movie S1).
The convex domains of the state $s=2$ grow toward their stable size; subsequently, they change to the $s=1$ state, and the resultant $s=1$ domains shrink, because the $s=1$ domains are unstable compared with the unbound membranes.
For example, in the snapshot taken at $t=50$ shown in Fig.~\ref{fig:BD}(a), green ($s=1$) domains are growing in four red $s=2$ domains.
These green domains shrink from $t=54$ to $66$.
This blinking behavior repeats, as shown in Movie S1.
These cyclic changes arise from chemical potential differences as in the three-state active Potts lattice model.\cite{nogu24a,nogu24b}
However, domain boundaries propagate in these lattice models.
In contrast, convex membrane shapes maintain the domain locations in BD. When the nucleation of a domain starts before the relaxation of convex shapes,
the domain is formed in almost the same location. 
Convex membranes (high $z$) exhibit cyclic changes of states (blinking)
and dimpled membranes (low $z$) exhibit few changes, as shown in Fig.~\ref{fig:BD}(b) and (c).
Therefore, the BD pattern is a stationary wave,
and the membrane deformation act as a memory of the previous pattern.

Between BD and the $s=1$-state dominant phase E$_1$, flat domain (FD) modes emerge (indicated by cyan diamonds in the phase diagram in Fig.~\ref{fig:pdt1}).
In the FD modes, amoeba-like or elongated domains move diffusively,
since they do not have asymmetric structures in a front-to-rear direction (see Figs.~\ref{fig:FD}, S2, and S3 and Movie S2). 
The dominant and second major states depend on the condition ($s=1$ and $2$ in Fig.~\ref{fig:FD}, $s=2$ and $1$ in Fig.~S2, and $s=1$ and $0$ in Fig.~S3, respectively). The third state exists to a small degree at the domain boundaries.
When proteins bind to both membrane surfaces,
a similar diffusive motion occurs for domains without front-to-rear asymmetry,  whereas front-to-rear asymmetric biphasic domains move balistically.\cite{nogu25a}

To quantify these dynamics, we calculate the mean density $\langle\phi_{s}\rangle$, 
 cluster size ratio $\chi_{s}$, mean vertical membrane span 
$\langle {z_{\mathrm {mb}}}^2\rangle$, and mean flipping rate $q_{\mathrm{f}}$ between neighboring states (see Fig.~\ref{fig:u0u1}).
In the E$_s$ and HD phases, these quantities exhibit behaviors similar to those in the equilibrium (compare  Fig.~\ref{fig:u0u1}(a), (c) and Fig.~\ref{fig:eq}(c), (d)).
In the HD and BD modes, the $s=0$ state forms a percolated network, i.e.,  $\chi_0\approx 1$.
In the FD mode, the domain of the $s=1$ or $s=2$ state is percolated ($\chi_1\approx 1$ or $\chi_2\approx 1$),
and the membrane is flat ($\langle {z_{\mathrm {mb}}}^2\rangle \simeq 0$).

The HD phase can continuously change to the BD mode, as shown in Fig.~\ref{fig:u0u1}(a)--(d).
To distinguish these modes, we calculate the time fraction $f_{\mathrm{3co}}$ of the three-state coexistence.
Figure \ref{fig:BD2} shows the BD pattern near the mode boundary,
in which $s=1$ domains occasionally appear in the hexagonal $s=2$ domains.
In this study, the pattern is determined as BD if $f_{\mathrm{3co}}>0.2$, where
the three states coexist when $\phi_{k}> \phi_{\mathrm {th}}$ for all states ($k\in {0,1,2}$).
The threshold $\phi_{\mathrm {th}}=0.005$ is used for most conditions, including that in Fig.~\ref{fig:BD2},
whereas $\phi_{\mathrm {th}}=0.01$--$0.03$ is used when the third state exists to a significant degree at the domain boundaries.

As $\mu_{02}$ increases in the BD mode at  $\mu_{01}=1$,
the flipping rate $q_{\mathrm{f}}$ and $\langle\phi_{1}\rangle$ increase, indicating that the blinking occurs more frequently (see Fig.~\ref{fig:u0u1}(e) and (h)). 
Subsequently, the $s=1$ domain becomes percolated, and the FD mode emerges.
Note that the entropy production rate\cite{herp18,agra25} for the steady state  is $(\mu_{02}-\mu_{01}-\mu_{12})q_{\mathrm{f}}$.\cite{nogu26c}

We have described the dynamic modes at $\mu_{12}=0$ up to this point.
A decrease in $\mu_{12}$ has similar effects to those of an increase in $\mu_{01}$, as shown in Fig.~S4.
The phase behaviors at $\mu_{12}=-1$ and  $\mu_{01}=0$ are shown in Fig.~S5.
These are very similar to those at $\mu_{12}=0$ and  $\mu_{01}=1$ (compare Fig.~S5 and Fig.~\ref{fig:u0u1}(e)--(h)).
Therefore, we consider that the essential dynamics are captured by the phase diagram at $\mu_{12}=0$.

\subsection{Nonequilibrium Patterns at $\gamma=0$ and $0.5$}\label{sec:t0}

The tensionless membranes ($\gamma=0$) exhibit no stable HD domains,  in either equilibrium or nonequilibrium.
However, the BD mode emerges at $\mu_{01}\simeq 2$ and $\mu_{02}\simeq 8$ as shown in the dynamic phase diagram of Fig.~\ref{fig:pdt0}.
The blinking convex domains shrink before growing to spherical buds,
resulting in the domains exhibiting steady oscillations (see Fig.~\ref{fig:BDt0} and Movie S3).
However, at too small or large $\mu_{02}$, vesicles are formed via spherical buds as shown in Fig.~\ref{fig:VESt0}. In the region of the black asterisk in  Fig.~\ref{fig:pdt0}, vesiculation occurs from any of the initial membrane conditions.
Therefore,  BD is a robust dynamic mode that can emerge even at $\gamma=0$, wherein the HD phase is inevitably unstable.

At $\gamma=0.5$, the membranes exhibit intermediate behaviors between those at $\gamma=0$ and $\gamma=1$,
as shown in Fig.~S6.
The BD mode and vesiculation emerge at $\mu_{01}\simeq 2$ and $\mu_{02}\simeq 8$
and at $\mu_{01}\simeq 0$ and $\mu_{02}\simeq 7$, respectively.
The HD phase emerges at $\mu_{01}\simeq 4$ and $\mu_{02}\simeq 8$, in contrast to the $\gamma=0$ dynamics.

In this study, we fixed the time units of state-flip and membrane dynamics ($\tau_{\mathrm {MC}}$ and $\tau$).
The ratio of the state-flip and membrane shape relaxation rates can be varied by the time ratio $\tau_{\mathrm {MC}}/\tau$.
When a much slower flip rate is used,
the blinking patterns will disappear owing to a loss of spatial memory.

\section{Summary}\label{sec:sum}

We have studied the spatiotemporal patterns of the membranes induced by protein binding and unbinding.
The proteins have two states that cause two different spontaneous curvatures in bound membranes.
In thermal equilibrium, the bound proteins can form hexagonally aligned convex domains owing to the competition of bending energy and surface tension.
In nonequilibrium, the convex domains can exhibit blinking, wherein the domains repeat growth and shrinkage.
The membrane bending maintains these domains around the original locations, as in standing waves.
Unlike the equilibrium convex domains, domain blinking emerges even in tensionless membranes.
Therefore, it is a robust pattern on deformable membranes
for wide ranges of chemical potentials and surface tensions.

In the present and previous studies,\cite{nogu25a} we used a single protein type with two states.
In general, proteins can have more than two states, and several protein types can cooperatively bind to membranes.
In the previous simulations of the $q$-state active Potts lattice models, we obtained patterns of factorized symmetry\cite{nogu25b,nogu26c} and competition among different cycles\cite{nogu26a} for $q>3$.
The coupling of these pattern dynamics and membrane deformation is an interesting topic for further studies.

Surface deformation occurs in lipid and cell membranes as well as in other chemical reaction systems.
The traveling waves of Belousov--Zhabotinsky and other reactions can accompany the shape oscillations of gel sheets.\cite{yosh22,mall25,livn24,tang26} 
On catalytic surfaces, such as noble metal surfaces such as palladium and ruthenium,
chemical waves have been observed experimentally (NO$_2$ reduction and CO oxidation, etc.).\cite{ertl08,bar94,goro94,scha02,barr20}
These waves can be accompanied by shape deformation of metal particles.\cite{tang20,ghos22}
In such systems, the surface deformation may alter the spatiotemporal patterns. Curved surfaces may pin or modify the wave propagation,
as observed in our study.

\begin{acknowledgments}
The simulations were partially carried out at ISSP Supercomputer Center, University of Tokyo (ISSPkyodo-SC-2025-Ca-0049).
This work was supported by JSPS KAKENHI Grant Number JP24K06973. 
\end{acknowledgments}


%

\end{document}